\documentclass{article}
\usepackage{graphicx} 
\usepackage{amsmath}
\usepackage[affil-it]{authblk}
\usepackage{mathtools}
\usepackage{geometry}
\newgeometry{left=1cm,right=0.1cm,bottom=3cm}

\title{SU(N) algebras and new thumbrules for entanglement of bipartite qubit and qutrit systems }

\author{P. Dasgupta \and D. Gangopadhyay }

\affil{ Department of Physics, Sister Nivedita University, DG Block , Newtown, Action Area 1, 
	
	Kolkata 700156, India }

\begin{document}
	\maketitle
	
	\abstract{Based on the Schmidt decomposition new convenient thumbrules are obtained to test entanglement of wavefunctions for bipartite qubit and qutrit systems. For the qubit system there is an underlying SU(2) algebra , while the same for a qutrit system is SU(3).} 
	
	\section{Introduction} 
	The concept of entanglement  materialised first in the papers [1-3]. Quantum 
	entanglement is associated with  nonclassical correlations  between spatially separated quantum 
	systems. One of the fundamental entities in the quantum domain is the wavefunction from which all physical measurables of a quantum 
	system may be obtained. A way to technically understand entanglement at the level of wavefunctions is that the composite quantum system wavefunction cannot be written as a direct product of the wavefunctions of the subsystems.
	
	There has been considerable literature regarding various aspects of entanglement of wave functions [10-32]. In this letter we discuss two new results related to entanglement of wavefunctions for bipartite qubit and qutrit systems. These are (a) there exist simple thumbrules to test entanglement based on the Schmidt decomposition for bipartite qubit and qutrit systems and (b) there is an underlying special unitary symmetry algebra in both cases.The plan of the letter is as follows: Section 2 is a quick review of the basic entities . Sections 3 and 4 contain the new results for bipartite qubit states while Section 5 contains our results for bipartite qutrit systems. Section 6 comprises our conclusions.
	\section{Qubits, Density matrix, Entanglement 
		and Bell states} : 
	We first give a quick review of some basic entities and concepts [5-9].
	In classical computation information is stored in a "bit" i.e 0 or 1. In quantum computation information is stored in quantum bits or qubits which can be $|0\rangle$, $|1\rangle$, or a superposition of both $|0\rangle$ and $|1\rangle$. A general qubit is:
	$|\psi\rangle=\alpha |0\rangle+\beta|1\rangle$,where $|\alpha|^2$ and $|\beta|^2$ are the probabilities to find $|\psi\rangle$ in states $|0\rangle $ and $ |1\rangle$ 
	respectively. $\alpha,\beta$ are the respective amplitudes and are complex scalars in general. The computational basis comprise 
	$|0\rangle=\begin{pmatrix}
		1 \\
		0
	\end{pmatrix}
	$
	$|1\rangle=\begin{pmatrix}
		0 \\
		
		1
	\end{pmatrix}$
	
	For a bipartite system , if  $\mathcal H_{A}$  and $\mathcal H_{B}$ are the Hilbert spaces of the two parts with 
	$\{|i\rangle_A\}$,
	$\{|\mu\rangle_B\}$ the orthonormal basis respectively, 
	then $\{|i\rangle_A \otimes |\mu\rangle_B \}$ is the orthonormal basis in $\mathcal H_A \otimes \mathcal H_B$. So 
	an arbitrary bipartite state in $\mathcal H_A \otimes \mathcal H_B$ can be written as
	$|\psi\rangle_{AB}= \Sigma_{\mu,i}    a_{i\mu}|i\rangle_A \otimes |\mu\rangle_B \equiv a_{i\mu}|i\rangle_A \otimes |\mu\rangle_B $, where we follow the Einstein convention, that is, summation over repeated indices.
	
	The density matrix of a given state $|\psi\rangle$ is
	$\rho=|\psi\rangle\langle\psi|$ and 
	the expectation value of an operator $M$ is
	$\langle M\rangle=tr(M\rho)$.
	The expectation value of an operator $M_A$ acting on subsystem $A$ is
	$\langle M_A\rangle = _{AB}\langle\psi|M_A \otimes I_B|\psi\rangle_{AB}=tr(M_A \rho_A)$. 
	Here 
	$\rho_ A  = tr_{B}(|\psi\rangle_{AB}.  _{AB}\langle\psi|)=tr_B(\rho_{AB}) 
	= _{B}\langle \mu|\rho_{AB}|\mu\rangle_{B} = a_{i\mu} a_{j\mu}^*|i\rangle_{A}. _A \langle j|$.
	$\rho_A$ is the reduced density matrix obtained by partial trace over sub system B.
	In a similar manner one has
	$\rho_B=tr_A(\rho_{AB})=  _{A}\langle i|\rho_{AB}|\i\rangle_{A}$.
	
	For the bipartite system if  the Hilbert space of the  composite system  $\mathcal H_{AB}\neq\mathcal H_A \otimes \mathcal H_B$, then the composite  state
	$|\psi\rangle_{AB}\neq|\psi\rangle_A\otimes|\psi\rangle_B$ and   $|\psi\rangle_{AB}$  is  said to be entangled.
	
	A composite state is said to be maximally entangled when the correlation between the subsystems is  maximum. 
	If $dim(\mathcal H_A)=dim(\mathcal H_B)=d$, then for maximal entanglement the reduced density matrices satisfy  $\rho_A=\rho_B=\frac{1}{d}I$ where 
	$I$ is the identity operator.
	
	The Bell States form the simplest example of maximally entangled states. They  also  form an orthonormal basis for a bipartite entangled state. The Bell States are
	\begin{equation}
		|\phi_0\rangle = \frac{|00\rangle+|11\rangle}{\sqrt{2}} ,
		\enskip
		|\phi_1\rangle=\frac{|01\rangle+|10\rangle}{\sqrt{2}} ,|\phi_{2}\rangle=\frac{|01\rangle-|10\rangle}{\sqrt{2}},
		|\phi_3\rangle=\frac{|00\rangle-|11\rangle}{\sqrt{2}}
	\end{equation}
	with $\langle\phi_i|\phi_j\rangle=\delta_{ij}$
	
	For the bipartite state $|\psi\rangle$  there exist othonormal states $|j_A\rangle,|j_B\rangle$ for system $A$ and $B$ respectively, such that
	\begin{equation}
		|\psi\rangle=\sum_j\sqrt{\mu_j}|j_A\rangle|j_B\rangle
	\end{equation}
	where $\mu_i$ are non-negative real numbers such that,
	$\sum_i\mu_i=1$. The above expression is known as the Schmidt decomposition [4]. Schmidt number is the number of non-zero eigen values of the reduced density matrix. If the number of terms in Schmidt decomposition series (equation (2)) is unity, i.e. the number of  non-zero eigenvalues of reduced density matrix is one, then the state is not entangled.
	If the number of terms exceeds unity then $|\psi\rangle$ is said to be an entangled state.
	
	The reduced density matrices are
	$\rho_A=\sum_i\mu_i|i_A\rangle\langle i_A|$, 
	and $\rho_B=\sum_i\mu_i|i_B\rangle\langle i_B|$
	and the eigenvalues of both the density matrices are same and equal to $\mu_i$.
	Peres {\it et al} gave a necessary condition criterion for separability [21,22]. 
	
	\section{ Bipartite states and entanglement for qubits :}
	We now determine the new results of this letter for bipartite qubit staes.
	
	Let us take the 2-state systems A and B  with Hilbert spaces $\mathcal H_A$ and $\mathcal H_B$ and basis vectors 
	$(|0\rangle_A , |1\rangle_A)$ 
	and 
	$(|0\rangle_B ,|1\rangle_B)$respectively.
	We work in the computational basis. 
	Thus, the basis vectors in the bipartite state AB are:
	$|00\rangle,|01\rangle,|10\rangle,|11\rangle$. The first entry in the ket denotes system A while the second entry denotes system B  and
	$\langle ij|i'j'\rangle=\delta_{i,i'}\delta_{j,j'}$
	where $(i,j)$ is $0$ or $1$.
	
	The most general bipartite state can be written as:
	\begin{equation}
		|\psi\rangle=a_{00}|00\rangle+a_{01}|01\rangle+a_{10}|10\rangle+a_{11}|11\rangle 
	\end{equation}
	Since we take normalized states,
	$ a_{00}^2+a_{01}^2+a_{10}^2+a_{11}^2=1$.
	
	Consider the matrix $A$ constructed as
	\begin{equation}
		A=\begin{pmatrix}
			a_{00} && a_{01}\\ 
			
			a_{10} && a_{11}
		\end{pmatrix}
	\end{equation}
	For simplicity we take $a_{ij}$ to be real.
	
	Now consider the Schmidt decomposition of the state using equation (2).
	
	The density matrix is 
	$\rho=|\psi\rangle\langle\psi|$ and 
	reduced density matrix is 
	\begin{equation} 
		\rho_A = Tr_B(\rho) = _B\langle0|\psi\rangle\langle\psi|0\rangle_B+ _B\langle1|\psi\rangle\langle\psi|1\rangle_B
		=\begin{pmatrix}
			(a_{00}^2+a_{01}^2) && (a_{00}a_{10}+a_{01}a_{11})\\
			
			(a_{00}a_{10}+a_{01}a_{11}) && (a_{11}^2+a_{10}^2)
		\end{pmatrix}
	\end{equation}
	
	The  reduecd density matrix $\rho_A$ is a 2x2 matrix thus it will have 2 eigen values which are $\mu_{i}$,(i=1,2) :
	\begin{equation}
		\mu_1 = (1/\sqrt{2}) (a_{00}^2 + a_{01}^2  a_{10}^2  +a_{11}^2 -(a_{00}^2 + a_{01}^2 + a_{10}^2 + a_{11}^2)^2-4(a_{01}^2a_{10}^2 - 2a_{00}a_{10}a_{01}a_{11} + a_{00}^2a_{11}^2)^{1/2})
		=(1/\sqrt{2})(1-\sqrt{1-4(DetA)^2})
	\end{equation}
	Similary,the other eigenvalue is
	\begin{equation}
		\mu_2=(1/\sqrt{2})(1+\sqrt{(1-4(DetA)^2}
	\end{equation}
	Thus we have the eigenvectors $|\mu_1\rangle$ and $|\mu_2\rangle$ whose exact mathematical forms are not needed to find the Schmidt number.
	
	Thus we get the Schmidt Decomposition series:
	\begin{equation}
		\label{S.D.S}
		|\psi\rangle=\sqrt{\mu_1}|\mu_1\rangle|\mu_1\rangle+\sqrt{\mu_2}|\mu_2\rangle|\mu_2\rangle 
	\end{equation}
	Clearly if $DetA =0$ then, $\mu_1=0$,$\mu_2 \neq 0$, the Schmidt decomposition series  (\ref{S.D.S}) becomes
	\begin{equation}
		|\psi\rangle=\sqrt{\mu_2}|\mu_2\rangle|\mu_2\rangle
	\end{equation}
	In equation (8) there is only one term and the Schmidt number is unity. Thus if $DetA=0$ the state is {\it not} entngled.
	Thus the condition for entanglement in a general bipartite basis is $DetA \neq 0$.

	{\it Range of Det A and Maximal Violation :}
	
	The state $|\psi\rangle$ is maximally entangled,if
	\begin{equation}
		\rho_A=
		\begin{pmatrix}
			a_{00}^2+a_{01}^2 && a_{00}a_{10}+a_{01}a_{11}\\
			
			a_{00}a_{10}+a_{01}a_{11} && a_{11}^2+a_{10}^2 \\
		\end{pmatrix}
		=\frac{I}{2}
	\end{equation}
	From above equation we get the range of $DetA$ i.e
	\begin{equation}
		0 \leq|DetA|\leq \frac{1}{2}
		\label{myeq}
	\end{equation}
	It can be shown that similar results can be also demonstrated for $\rho_B$.
	Thus for maximal violation $|DetA|< \frac{1}{2}$ and greater the value of $|DetA|$, the more entangled the state is.
	
	{\it Example:}
	
	Consider the situation discussed in [30]. The state is $|\psi\rangle = cos\theta |00\rangle + sin\theta |11\rangle$ and  here $DetA=sin (2\theta)/2$. When  
	$DetA = sin (2\theta)/2 = 1/2$  the state is  maximally entangled for  $\theta=n\pi/4$ and not entangled for $DetA=sin (2\theta)/2=0$, that is for $\theta=n\pi/2$. 
	
	\section{The Qubit State in Bell Basis:}
	
	The most general  state $|\eta\rangle $ can be written as a superposition of Bell states also.
	\begin{equation}
		|\eta\rangle=b_0|\phi_0\rangle+b_1|\phi_1\rangle+b_2|\phi_2\rangle+b_3|\phi_3\rangle
	\end{equation}
	where, $\langle\eta|\eta\rangle=1$ ,
	i.e. $b_0^2+b_1^2+b_2^2+b_3^2=1$
	
	Plugging in the expressions for the Bell states from equation (1) one gets
	
	\begin{equation}
		|\eta\rangle = (1/2)^{1/2} [ (b_0+b_3)|00\rangle 
		+ (b_1+b_2)|01\rangle \enskip\enskip\enskip 
		+  (b_1 - b_2)|10\rangle +(b_0-b_3)|11\rangle] \enskip\enskip\enskip
		\enskip\enskip\enskip\enskip\enskip\enskip\enskip\enskip\enskip\enskip\enskip\enskip\enskip\enskip\enskip
	\end{equation}
	
	Note that under the identification of 
	
	$a_{00}= (1/2)^{1/2}(b_0+b_3) , a_{01}= (1/2)^{1/2}(b_1+b_2) ,  
	a_{10}= (1/2)^{1/2}(b_1-b_2) , a_{11}= (1/2)^{1/2}(b_0-b_3)$ , 
	
	we get back equation (3) and $|\eta\rangle \equiv |\psi\rangle $. The normalisation conditions are also  consistent with each other.
	
	We now discuss the Schmidt decomposition for $|\eta\rangle$.
	First we define the matrix,
	\begin{equation}
		C= \frac{1}{\sqrt{2}}\begin{pmatrix}
			b_0+b_3 && b_1+b_2\\
			b_1-b_2 && b_0-b_3
		\end{pmatrix}
	\end{equation}
	Here $b_i$'s are real.The density matrix of $|\eta\rangle$, $\rho=|\eta\rangle\langle\eta|$ and the reduced density matrix corresponding to subsystem A is 
	\begin{equation}
		\rho_A=tr_B(|\eta\rangle\langle\eta|)= _B\langle0|\eta\rangle\langle\eta|0\rangle_B +  _B\langle1|\eta\rangle\langle\eta|1\rangle_B 
	\end{equation}
	As we are dealing with a 2x2 matrix, $\rho_A$ has two eigenvalues $\mu_1$ and $\mu_2$.
	\begin{equation}
		\begin{split}
				\mu_1=(1/2)((b_0^2+b_1^2+b_2^2+b_3^2)-((b_0^2+b_1^2+b_2^2+b_3^2)^2-((b_1^2-b_2^2)^2
			-2(b_1^2-b_2^2)(b_0^2-b_3^2)+(b_0^2-b_3^2)^2))^{1/2})\\=(1/2)(1-\sqrt{1-((b_0^2-b_3^2)-(b_1^2-b_2^2))^2}=(1/2)(1-\sqrt{1 -(DetC)^2} 
					\end{split}	
	\end{equation}
	
	Similarly,
	\begin{equation}
		\mu_2=(1/2)(1+\sqrt{1-(DetC)^2}\enskip\enskip\enskip
	\end{equation}
	Thus the Schmidt Decomposition Series will be
	\begin{equation}
		|\eta\rangle=\sqrt{\mu_1}|\mu_1\rangle|\mu_1\rangle+\sqrt{\mu_2}|\mu_2\rangle|\mu_2\rangle
	\end{equation}
	
	But if $DetC=0$ then $\mu_1=0$ and the Schmidt Decomposition series has only one term
	
	$|\psi\rangle=\sqrt{\mu_2}|\mu_2\rangle|\mu_2\rangle$
	and  the state is not entangled.
	
	Thus the condition for not entanglement in Bell basis is  $ DetC=0$ which means:
	\begin{equation}
		b_0^2 + b_2^2 = b_1^2+b_3^2
	\end{equation}
	This means 
	\begin{equation}
		\langle\phi_0|\eta\rangle^2+\langle\phi_2|\eta\rangle^2=\langle\phi_1|\eta\rangle^2+\langle\phi_3|\eta\rangle^2
	\end{equation}
	
	Thus we can say that for every basis system we can form unique matrices whose determinant can be calculated and if the determinant is zero, we say the state is not entangled. If the determinant is non-zero the state  
	is entangled.
	
	{\it The A matrix for Bell States :}
	
	Now we construct the $A$ matrix for  the four Bell States and show that there exists an underlying $SU(2)$ algebra. 
	The general state can be written as (shown previously)
	$|\psi\rangle=a_{00}|00\rangle+a_{01}|01\rangle+a_{10}|10\rangle+a_{11}|11\rangle$
	
	The A matrix for the above state is
	$$A=\begin{pmatrix}
		a_{00} && a_{01}\\
		
		 a_{10} && a_{11}
		\end{pmatrix}$$
	
	When
	 $|\psi\rangle= |\phi_0 \rangle$
	$a_{00}=\langle 00|\phi_0 \rangle=1/\sqrt{2}$,
	$a_{01}=\langle 01|\phi_0 \rangle=0$, 
	$a_{10}=\langle 10|\phi_0 \rangle=0$, 
	$a_{11}=\langle 11|\phi_0 \rangle=1/\sqrt{2}$
	so that
	\begin{equation}
		A_0=\frac{1}{\sqrt{2}}\begin{pmatrix}
			1 && 0\\
			
			 0 &&1
\end{pmatrix}			 
		=\frac{I}{\sqrt{2}}
	\end{equation}
	
	Similarly, for $|\psi\rangle= |\phi_1 \rangle$
	\begin{equation}
		A_1=\frac{1}{\sqrt{2}}\begin{pmatrix}
			0 && 1\\
			
			 1 &&0
			 \end{pmatrix}
		=\frac{1}{\sqrt{2}}\sigma_1
	\end{equation}
	For $|\psi\rangle= |\phi_2 \rangle$
	\begin{equation}
		A_2=\frac{1}{\sqrt{2}}\begin{pmatrix}
			0 && 1\\
			
			 -1 &&0
			\end{pmatrix}
		=\frac{i}{\sqrt{2}}\sigma_2
	\end{equation}
	For $|\psi\rangle= |\phi_3 \rangle$
	\begin{equation}
		A_3=\frac{1}{\sqrt{2}}\begin{pmatrix}
			1 && 0\\
			
			 0 &&-1
			\end{pmatrix}
		=\frac{1}{\sqrt{2}}\sigma_3
	\end{equation}
	where $I$ is the unit matrix and $\sigma_i$
	are Pauli Matrices.
	Redefining variables :
	$A_1'=\sqrt{2}A_1$ , $A_2'=\sqrt{2}e^{(-i\pi)/2}A_2$, $A_3'=\sqrt{2}A_3$, 
	one obtains the $SU(2)$ algebra
	\begin{equation}
		\label{SU2}
		[A_i',A_j']=2i\epsilon_{ijk}A_k'
	\end{equation}
	
	The commutation relations for $A_i$ matrices are:
	\begin{equation}
		[A_1,A_2]=-\sqrt{2}A_3,\enskip
		[A_2,A_3]=-\sqrt{2}A_1,\enskip
		[A_3,A_1]=\sqrt{2}A_2
	\end{equation}
	Now, $\sigma_2=\sigma_2^{\dagger}$. Therefore 
	$A_2 = -A_2^{\dagger}$. Thus $A_2$ is pure imaginary.
	So one may write $A_2\equiv K^{\dagger}U$, 
	where $K^{\dagger}$ is complex conjugation and $U$ is a unitary operator. Similar scenarios have  been reported elsewhere [33] in a totally different context.
	
	\section{The bipartite qutrit system:}
	The simplest 3-state quantum system is called the qutrit. The most general qutrit is $|\mu\rangle=\alpha|0\rangle+\beta|1\rangle+\gamma|2\rangle$, where
	the column vectors 
	$|0\rangle=\begin{pmatrix}
		1 && 0 && 0
		\end{pmatrix}^T\enskip $\  ;
	$|1\rangle=\begin{pmatrix}
		0 && 1 && 0
	\end{pmatrix}^T\enskip $
	$|2\rangle=\begin{pmatrix}
		0 && 0 && 1
	\end{pmatrix}^T\enskip $
	 form an orthonormal basis.
	Let there be two systems A and B belonging to 3D Hilbert spaces $\mathcal H_{A}$ and  $\mathcal H_{B}$ respectively. The basis vectors for the composite system AB belonging to the 9D Hilbert space, $\mathcal{H_{AB}}$ are $\{ |00\rangle , |01\rangle . |02\rangle , |10\rangle , |11\rangle , |12\rangle , |20\rangle , |21\rangle , |22\rangle \}$ , where $\langle i i'| j j'\rangle = \delta_{ij}\delta_{i'j'}$ , with $i,j=0,1,2$.The most general quantum state in the composite system AB,
	
	\begin{equation}
		|\chi\rangle  
		= a_{00}|00\rangle + a_{01}|01\rangle + a_{02}|02\rangle 
		+ a_{10}|10\rangle + a_{11}|11\rangle + a_{12}|12\rangle
		+ a_{20}|20\rangle + a_{21}|21\rangle + a_{22}|22\rangle 
	\end{equation}
	
	Define the matrix,
	\begin{equation}
		P=\begin{pmatrix}
			a_{00} && a_{01} && a_{02}\\
			
			 a_{10} && a_{11} && a_{12}\\
			  a_{20} && a_{21} && a_{22}
			\end{pmatrix}
	\end{equation}
	The density matrix for the state $|\chi \rangle_{AB}$ is
	$\rho_{AB}=|\chi\rangle_{AB} ._{AB}\langle\chi|$ and 
	the reduced density matrix is
	\begin{equation}
		\rho_A=Tr_b(\rho_{AB})=\langle 0|\rho_{AB}|0\rangle+\langle 
		1|\rho_{AB}|1\rangle+\langle 2|\rho_{AB}|2\rangle = PP^{\dagger}\enskip\enskip\enskip  
	\end{equation}
	Thus, $Det\rho_A=DetPDetP^{\dagger}$. Hence $|Det\rho_A|=|DetP|^2$.
	
	$P$ is a 3x3 matrix and it will have three eigenvalues $\mu_1, \mu_2, \mu_3$. Then the eigenvalues of $\rho_A=PP^{\dagger}$ are $|\mu_1|^2,|\mu_2|^2 and |\mu_3|^2$.
	
	For normalization
	\begin{equation}   
		\label{Normalization}
		Tr\rho_A=|\mu_1|^2+|\mu_2|^2+|\mu_3|^2 =1
	\end{equation}
	For invariance of trace
	\begin{equation}
		\label{Invariance of trace}
		TrP=\mu_1+\mu_2+\mu_3
	\end{equation}
	For invariance of determinant
	\begin{equation}
		Det\rho_A=|DetP|^2=\mu_1^2.\mu_2^2.\mu_3^2
	\end{equation}
	So
	\begin{equation}
		DetP=\mu_1.\mu_2.\mu_3
	\end{equation}
	The Schmidt decomposition series for $\rho_A$ is
	\begin{equation}
		\label{qutrit schmidt}
		|\chi\rangle_{AB}=\mu_1|\mu_1\rangle|\mu_1\rangle+\mu_2|\mu_2\rangle|\mu_2\rangle+\mu_3|\mu_3\rangle|\mu_3\rangle
	\end{equation}
	In general $DetP \neq 0 => \mu_1\mu_2\mu_3 \neq 0$. Therefore 
	$\mu_i\neq 0$ i=1,2,3 . So there are three terms in  equation (\ref{qutrit schmidt})which means that the state $|\chi\rangle_{AB} $ is entangled.
	
	For the state to be unentangled,the series must have only one term,so we start with the condition
	\begin{equation}
		\label{det0}
		DetP=0=>\mu_1\mu_2\mu_3 =0 
	\end{equation}
	This means either one or two of the eigenvalues must be zero.
	Solving equations (\ref{Normalization}),(\ref{Invariance of trace}),(\ref{det0}), we get
	
	\textbf{Case 1:}
	$\mu_1=0 ,\mu_2=\frac{1}{2}(TrP-\sqrt{2-TrP^2}) ,\mu_3=\frac{1}{2}(TrP+\sqrt{2-TrP^2})$
	
	Thus the Schmidt decomposition series (\ref{qutrit schmidt}) reduces to
	$|\chi\rangle_{AB}=\mu_2|\mu_2\rangle|\mu_2\rangle+\mu_3|\mu_3\rangle|\mu_3\rangle$. If $TrP=+1$ then $\mu_2=0 , \mu_3=1$ and the above series reduces further into $|\chi\rangle_{AB}=|\mu_3\rangle|\mu_3\rangle$ which has only one term. Thus the state is not entangled.
	
	If $TrP=-1$ then $\mu_3=0 , \mu_2=-1$ and the above series reduces further into $|\chi\rangle_{AB}=-|\mu_2\rangle|\mu_2\rangle$ which also has only one term and thus the state is not entangled.
	
	\textbf{Case 2:}
	$\mu_1= \frac{1}{2}(TrP-\sqrt{2-TrP^2}),\mu_2=0 ,\mu_3=\frac{1}{2}(TrP+\sqrt{2-TrP^2})$
	
	and 
	
	\textbf{Case 3:}
	$\mu_1= \frac{1}{2}(TrP-\sqrt{2-TrP^2}),\mu_2=\frac{1}{2}(TrP+\sqrt{2-TrP^2}),\mu_3=0$
	
	In both the cases 2 and 3, the conditions of a bipartite qutrit state to be unentangled  are, $DetP=0,TrP=\pm1$ as in case 1.
	
	\section{Qutrit entangled basis and Gell-mann matrices}
	The entangled bipartite qutrit basis are,
	\begin{equation}
		|\beta_0\rangle = \frac{|00\rangle+|11\rangle+|22\rangle}{\sqrt{3}} ,
		\enskip\enskip
		|\beta_1\rangle=\frac{|01\rangle+|10\rangle}{\sqrt{2}}
	\end{equation}
	\begin{equation}
		|\beta_2\rangle=\frac{|01\rangle-|10\rangle}{\sqrt{2}},
		\enskip\enskip
		|\beta_3\rangle=\frac{|00\rangle-|11\rangle}{\sqrt{2}} 
	\end{equation}
	\begin{equation}
		|\beta_4\rangle=\frac{|02\rangle+|20\rangle}{\sqrt{2}},
		\enskip\enskip
		|\beta_5\rangle=\frac{|02\rangle-|20\rangle}{\sqrt{2}}  
	\end{equation}
	\begin{equation}
		|\beta_6\rangle=\frac{|12\rangle+|21\rangle}{\sqrt{2}},
		\enskip\enskip
		|\beta_7\rangle=\frac{|12\rangle-|21\rangle}{\sqrt{2}}    
	\end{equation}
	\begin{equation}
		|\beta_8\rangle =\frac{|00\rangle+|11\rangle-2|22\rangle}{\sqrt{6}}
	\end{equation}
	where,$\langle \beta_i|\beta_j\rangle=\delta_{ij}$
	
	The Gell-Mann matrices are:
	\begin{equation}
		\lambda_1 = \begin{pmatrix}
			 0& 1 & 0 \\
			  
			  1 & 0 & 0 \\
			  
			   0 & 0 & 0
			\end{pmatrix},
		\lambda_2 = \begin{pmatrix}
			0 & -i & 0 \\ 
			
			i & 0 & 0 \\
			
			 0 & 0 & 0 
			\end{pmatrix}
	\end{equation}
	\begin{equation}
		\lambda_3 = \begin{pmatrix}
			1 & 0 & 0 \\
			
			 0 & -1 & 0 \\
			 
			  0 & 0 & 0 
			\end{pmatrix},
		\lambda_4 = \begin{pmatrix}
			
			 0 & 0 & 1 \\
			 
			  0 & 0 & 0 \\
			  
			   1 & 0 & 0
			\end{pmatrix}
	\end{equation}
	\begin{equation}
		\lambda_5 = \begin{pmatrix}
			 0 & 0 & -i \\
			 
			  0 & 0 & 0 \\
			  
			   i & 0 & 0 
			\end{pmatrix},
		\lambda_6 = \begin{pmatrix}
			 0 & 0 & 0 \\ 
			 
			 0 & 0 & 1 \\
			 
			  0 & 1 & 0 
			\end{pmatrix}
	\end{equation}
	\begin{equation}
		\lambda_7 = \begin{pmatrix}
			0 & 0 & 0 \\
			
			 0 & 0 & -i \\
			 
			  0 & i & 0
			\end{pmatrix},
		\lambda_8 = \frac{1}{\sqrt{3}} \begin{pmatrix}
			
			 1 & 0 & 0 \\
			  
			  0 & 1 & 0 \\
			  
			   0 & 0 & -2
			\end{pmatrix}
			\end{equation}
			
			Below are the values of the determinants and traces of all the $P_i$ matrice i= 0 to 8. Note that the determinant value is always understood in terms of its absolute value.
			
			If $|\chi\rangle_{AB}=|\beta_0\rangle=\frac{|00\rangle+|11\rangle+|22\rangle}{\sqrt{3}};a_{00}=a_{11}=a_{22}=\frac{1}{\sqrt{3}}$where.$a_{ij}=0,i\neq j$. 
			Thus 
			\begin{equation}
				\label{P0}
				P_0=\frac{1}{\sqrt{3}}I ,\enskip Det P_0 = 1/3\sqrt{3} ,\enskip Tr P_0 = \sqrt{3} \neq \pm 1
			\end{equation}
			
			If $|\chi\rangle_{AB}=|\beta_1\rangle=\frac{|01\rangle+|10\rangle}{\sqrt{2}};a_{01}=a_{10}=\frac{1}{\sqrt{2}};a_{ij}=a_{kl}=0$;where$i,j=2,0$ and $k,l=1,2$,thus
			\begin{equation}
				\label{P1}
				P_1=\frac{\lambda_1}{\sqrt{2}} ,\enskip Det P_1 = 0 ,\enskip Tr P_1 = 0 \neq \pm 1
			\end{equation}
			If $|\chi\rangle_{AB}=|\beta_2\rangle=\frac{|01\rangle-|10\rangle}{\sqrt{2}};a_{01}=-a_{10}=\frac{1}{\sqrt{2}};a_{ij}=a_{kl}=0$;where$i,j=2,0$ and $k,l=1,2$,thus
			\begin{equation}
				\label{P2}
				P_2=\frac{i \lambda_2}{\sqrt{2}} ,\enskip Det P_2 = 0 ,\enskip Tr P_2 = 0 \neq \pm 1
			\end{equation}
			If $|\chi\rangle_{AB}=|\beta_3\rangle=\frac{|00\rangle-|11\rangle}{\sqrt{2}};a_{00}=-a_{11}=1/\sqrt{2};a_{ij}=a_{kl}=0,a_{22}=0;i \ne j ; k \ne l$,where $i,j=0,1$ and $k,l=1,2$, thus
			\begin{equation}
				\label{P3}
				P_{3}=\frac{\lambda_3}{\sqrt{2}} , ,\enskip Det P_3 = 0 ,\enskip Tr P_3 = 0 \neq \pm 1
			\end{equation}
			If $|\chi\rangle_{AB}=|\beta_4\rangle=\frac{|02\rangle+|20\rangle}{\sqrt{2}};a_{02}=a_{20}=1/\sqrt{2};a_{ij}=a_{kl}=0$,where $i,j=0,1$ and $k,l=1,2$,thus
			\begin{equation}
				\label{P4}
				P_4=\frac{\lambda_4}{\sqrt{2}} ,\enskip Det P_4 = 0 ,\enskip Tr P_4 = 0 \neq \pm 1
			\end{equation}
			If $|\chi\rangle_{AB}=|\beta_5\rangle=\frac{|02\rangle-|20\rangle}{\sqrt{2}};a_{02}=-a_{20}=1/\sqrt{2};a_{ij}=a_{kl}=0$,where $i,j=0,1$ and $k,l=1,2$,thus
			\begin{equation}
				\label{P5}
				P_5=\frac{i \lambda_5}{\sqrt{2}} ,\enskip Det P_5 = 0 ,\enskip Tr P_5 = 0 \neq \pm 1
			\end{equation}
			If $|\chi\rangle_{AB}=|\beta_6\rangle=\frac{|12\rangle+|21\rangle}{\sqrt{2}};a_{12}=a_{21}=1/\sqrt{2};a_{ij}=a_{kl}=0$,where $i,j=0,1$ and $k,l=0,2$,thus
			\begin{equation}
				\label{P6}
				P_6=\frac{\lambda_6}{\sqrt{2}} ,\enskip Det P_6 = 0 ,\enskip Tr P_6 = 0 \neq \pm 1
			\end{equation}
			If $|\chi\rangle_{AB}=|\beta_7\rangle=\frac{|12\rangle-|21\rangle}{\sqrt{2}};a_{12}=-a_{21}=1/\sqrt{2};a_{ij}=a_{kl}=0$,where $i,j=0,1$ and $k,l=0,2$,thus
			\begin{equation}
				\label{P7}
				P_7=\frac{i \lambda_7}{\sqrt{2}} ,\enskip Det P_7 = 0 ,\enskip Tr P_7 = 0 \neq \pm 1
			\end{equation}
			If $|\chi\rangle_{AB}=|\beta_8\rangle=\frac{|00\rangle+|11\rangle-2|22\rangle}{\sqrt{6}};a_{00}=a_{11}=\frac{1}{\sqrt{6}};a_{22}=\frac{-2}{\sqrt{6}}$where.$a_{ij}=0,i\neq j$
			thus \begin{equation}
				\label{P8}
				P_8= \frac{\lambda_8}{\sqrt{2}} ,\enskip Det P_8 = 1/3\sqrt{6} ,\enskip Tr P_8 = 0 \neq \pm 1
			\end{equation}
			Redefining variables in equations (\ref{P1}),(\ref{P2}), (\ref{P3}),(\ref{P4}),(\ref{P5}),(\ref{P6}),(\ref{P7}),(\ref{P8}) we get
			\begin{equation}
				\begin{split}
					P_0'=\sqrt{3}P_0; P_1'=\sqrt{2}P_1; P_2'=\sqrt{2}e^{(-i\pi/2)}P_2 \\	P_3'=\sqrt{2}P_3; 
				P_4'=\sqrt{2}P_4;\\ P_5'=\sqrt{2}e^{(-i\pi/2)}P_5; 
				P_6'=\sqrt{2}P_6; \\
				P_7'=\sqrt{2}e^{(-i\pi/2)}P_7; P_8'= \sqrt{2} P_8 
				\enskip\enskip\enskip\enskip\enskip\enskip\enskip
				\end{split}			
			\end{equation}
			
			It is easily seen that,
			\begin{equation}
				[P_{\kappa}',P_{\mu}']=2i f_{\kappa\mu\nu}P_{\nu}'
			\end{equation}
			where  $\kappa,\mu,\nu=1,2,3,...,8$. Equation (55) represents a SU(3) algebra  and $f_{\kappa\mu\nu}$ are  the structure constants 
			of the SU(3) group.
			
			It is easily noted that $P_2=-P_2^{\dagger} $ ;$ P_5=-P_5^{\dagger}$ and $ P_7=-P_7^{\dagger}$. Thus we can write $P_i=K_i^{\dagger}U_i$ where $i=2,5,7$ where $K_i^{\dagger}$'s are complex conjugation operators and $U_i$'s are unitary operators. We have seen similar results in section 2. This is a new result in the context of qutrits.
			
			The reduced density matrix $\rho_A$  
			(or $\rho_B$) has 6 independent elements due to its symmetric nature. To find the maximally entangled states we equate this to $I/3$ as $\rho_A= P P^{\dagger} =\frac{I}{3}$  and thus we have 6 equations and 9 unkowns. 
			
			If $P= \pm P^{\dagger}$(i.e. hermitian or  anti-hermitian)  the number of independent variables will reduce to 6. Thus we have 6 equations and 6 unknowns which can be solved to get maximally entangled bipartrite qutrit states. For other forms of the matrix $P$ (unitary, orthogonal etc.) things are under investgation.
			
			\section{Conclusion:}
			In this letter we have developed simple thumbrules to test for entanglement of wavefunctions for bipartite qubit and bipartite qutrit systems. Our thumbrules are based on the Schmidt decomposition and are 
			easy to apply. 
			
			For the qubit system one can always construct a matrix $A$ (equation (4))using the corresponding amplitudes of the basis states. If $Det A = 0$, the state is unentangled. For $Det A \neq 0$ the state is entangled. We also demonstrate the existence of an underlying exact SU(2) algebra.
			This is demonstrated in equation (25) using the relevant constructions as defined in equations (21)-(24).
			
			For the bipartite qutrit system one can again construct a matrix $P$, 
			equation (28), using the amplitudes defined in equation (27). 
			The entangled bipartite qutrit basis states are taken as in equations (36)-(40).For the bipartite qutrit state along with the determinant P being zero, the trace of P matrix should also be $\pm 1$ in order for the state to be unentangled.For the entangled qutrit system a SU(3) closed algebra is obtained using a redefinition of quantities (equation (54)) using the Gell-Mann matrices (equations (41)-(44)).This is demonstrated in equation (55) using the relevant constructions as defined in equations (45)-(53). 
			
			Another interesting aspect follows from the discussions after equations (26) and (55). The underlying SU(2) or SU(3) algebras obtained in our work seems to be related to the  symmetry of time reversal in quantum systems. Recall that there is only one independent antiunitary symmetry whose physical meaning is  time reversal. Any other antiunitary transformation can be expressed in terms of
			time reversal (as the product of a unitary matrix times time reversal). This is well known and was discussed in [33].This fact has been built into our redefinition of the matrix $A_2$ as $A_ {2} '$ immediately after equation (24) for the qubit case. Similar redefinitions of $P_2 , P_5, P_7$ as $P_2 ', P_5 ', P_7 '$ respectively for the qutrit case have the same underlying physical meaning.
			
			The efficacy and efficiency in computing time using our method described in this work may be useful in recent exciting particle physics scenarios as described in the very recent report [34].
			\section{Acknowledgement}
			One of the author(PDG) would like to thank Sister Nivedita University for providing research scholarship, under Student ID-2331207003001

\end{document}